# Exploring Euroleague History using Basic Statistics

Christos Katris[1,2]

[1]*Adjunct Lecturer, Department of Mathematics, University of Patras*

[2]*Customs Officer (Statistician), Independent Authority for Public Revenue, Greece*

[1]chriskatris@upatras.gr, [2]c.katris1@aade.gr

**Abstract**

In this paper are used historical statistical data to track the evolution of the game in the European-wide top-tier level professional basketball club competition (until 2017-2018 season) and also are answered questions by analyzing them. The term basic is referred because of the nature of the data (not available detailed statistics) and of the level of aggregation (not disaggregation to individual level). We are examining themes such as the dominance per geographic area, the level of the competition in the game, the evolution of scoring pluralism and possessions in the finals, the effect of a top scorer in the performance of a team and the existence of unexpected outcomes in final fours. For each theme under consideration, available statistical data is specified and suitable statistical analysis is applied. The analysis allows us to handle and answer the above themes and interesting conclusions are drawn. This paper can be an example of statistical thinking in basketball problems by the means of using efficiently available statistical data.

## 1. Introduction

The field of basketball is ideal for the application of statistical methods in order to extract useful conclusions which can help in analyzing many aspects of the game. The origin of many ideas is from persons outside academia. The book of Oliver (2004) was a worthy attempt to develop and apply statistical concepts in the area of basketball. Much information is included on this book and can offer to a reader a statistical way of thinking for the game of basketball. There are also many academic papers which use advanced statistical methods for basketball analysis in themes such as performance evaluation of players and teams, home advantage effect etc. The field of basketball analytics is not yet entirely unified and new ideas which are based on quantitative analysis are appearing continuously from diverse academic fields. In many cases, there are used advanced statistics for the analysis of many situations. The majority of studies – not only with USA origin - are related to NBA and this is not just a coincidence. The tracking system of statistics is superior to other leagues in terms of quality (calculation of more advanced statistics) and quantity (calculation of more statistical categories), and the discrepancy was larger especially in the past.

The paper of Kubatko et al (2007) presents the general accepted basics of the analysis of basketball. Furthermore, most of the statistics are based on the concept of possessions. However, this is not the case for other leagues, including Euroleague. Given the available statistical data is difficult or even impossible (for older years) to calculate neither possessions nor advanced statistics. Only after 2001 in the modern era of Euroleague, plenty of statistics are available.

This paper is an attempt to utilize available statistical information through statistical analysis in order to explore the evolution of the game in Euroleague. It is demonstrated that even simple available statistics can offer insights about the game and can be extracted useful conclusions. Graphical analysis, statistical hypothesis testing and correlation measures are our weapons in this chase of insights related to the evolution of Euroleague. The next section is a brief description of Euroleague and are referred the sources of statistical data. Section 3 is the main part of the paper and contains statistical analysis and methods to deal with questions related to the historical evolution of the tournament. Finally, in Section 4 are presented the conclusions of the analysis.

## 2. A Brief History of Euroleague and Statistical Data

In this paper is examined the evolution of the European-wide top-tier level professional basketball club competition. Briefly the history of the competition is following. The FIBA European Champions Cup competition has established in 1958 and FIBA was organizing its operation until 2000. Then Euroleague Basketball was created. The next year, the two competitions were unified again under the umbrella of Euroleague Basketball (for more details: *https://en.wikipedia.org/wiki/EuroLeague*). Also the competition has changed names across time. From 1958 to 1991 was the *FIBA European Champions Cup*, from 1991 to 1996 the name of the competition was *FIBA European League*, from 1996 to 2000 the name was *FIBA*

*Euroleague*. In season 2000-2001 there were 2 competitions: *FIBA Suproleague* which was organized by FIBA and *Euroleague* which was organized by Euroleague Basketball. From the next year there was a unique competition for the top-tier level under the name *Euroleague* which was organized by Euroleague Basketball. In 2016 the name changed to *EuroLeague*. For the rest of the article the name Euroleague is used for the whole competition. The concept of final four applied for 1965-1966 and 1966-1967 seasons and was included permanently in the competition from the season of 1987-1988. In this paper, we consider as final four teams before 1986-1987, the teams which have reached the semi-finals in order to generate a consistent system for studying the evolution of the tournament.

There is not a unique data source which contains all information from the beginning of the tournament in 1958. Statistical data sources which were used are: *Wikipedia, http://pearlbasket.altervista.org, http://www.linguasport.com and http://www.fibaeurope.com/* and *http://www.euroleague.net/* for stats after 2001.

## 3. Statistical Analysis of Euroleague Historical Data

In this section is made an attempt to shed light to questions related to the historical evolution of the game with the use of suitable statistical methods. The graphs are created in excel, whilst for the implementation of the methods is used statistical software R.

### 3.1 Dominance on the Game per Geographic Area

Firstly, we can derive some quick conclusions about the dominance in the game in terms of geographic location. Table 1 displays per country the winners, the runners-up and the number of teams which had appeared to final fours. We consider only the teams which participated to final fours since 1958. From this limited statistical information we will explore briefly the game over time.

Based on Table 1, we consider the following Geographic areas: Spain and Italy which are leading the table in all categories are considered separately, Ex USSR and ex Yugoslavian countries form the next area and every other country is assigned to a fourth group (other).

Fig.1 displays the titles per time period of teams from each geographic area and Fig.2 displays the appearances in final fours of teams from each geographic area.

Table 1. Titles and appearances per country

| Country | Winner | Runner-Up | Final Four Appearances | Number of Teams |
|---|---|---|---|---|
| Spain | 13 | 16 | 57 | 6 |
| Italy | 13 | 13 | 44 | 9 |
| Greece | 9 | 7 | 29 | 5 |
| Russia[1] | 7 | 6 | 30 | 2 |
| Israel | 6 | 9 | 20 | 1 |
| Croatia[2] | 5 | 1 | 9 | 3 |
| Latvia[1] | 3 | 1 | 4 | 1 |
| Turkey | 1 | 2 | 6 | 2 |
| Lithuania[1] | 1 | 1 | 3 | 1 |
| Georgia[1] | 1 | 1 | 3 | 1 |
| Bosnia[2] | 1 | 0 | 4 | 1 |
| Serbia[2] | 1 | 0 | 10 | 4 |
| France | 1 | 0 | 9 | 4 |
| Czech Republic[3] | 0 | 3 | 9 | 2 |
| Bulgaria | 0 | 2 | 2 | 1 |
| Slovenia[2] | 0 | 0 | 3 | 1 |
| Poland | 0 | 0 | 2 | 2 |
| Romania | 0 | 0 | 1 | 1 |
| Netherlands | 0 | 0 | 1 | 1 |
| Hungary | 0 | 0 | 1 | 1 |
| Belgium | 0 | 0 | 1 | 1 |

[1] *Trophies won before 1991 were under the umbrella of Soviet Union*

[2] *Trophies won before 1995 were under the umbrella of Yugoslavia*

[3] *Trophies won before 1991 were under the umbrella of Czechoslovakia*

Fig.1. Titles evolution per geographic area

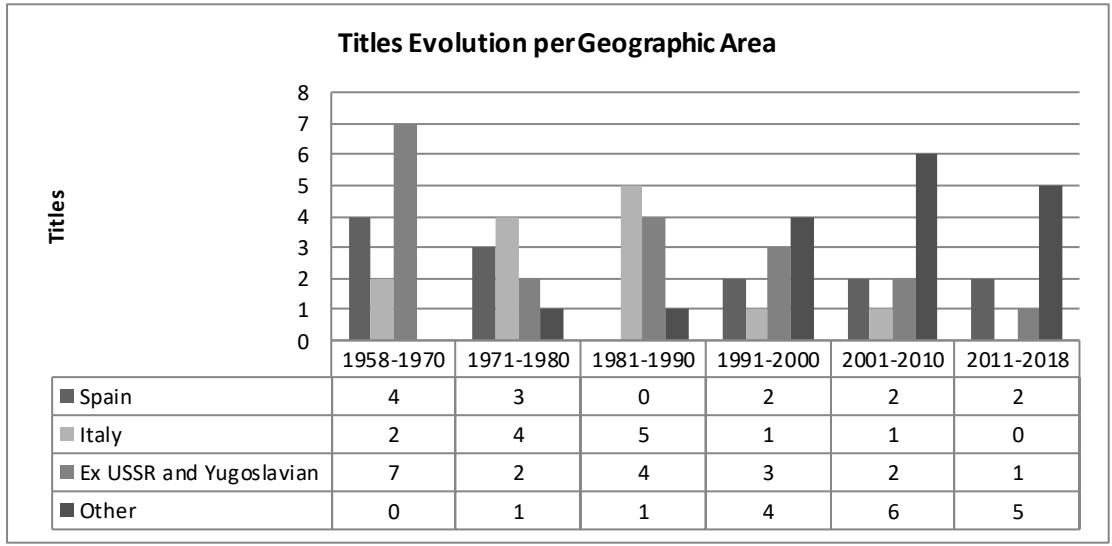

| | 1958-1970 | 1971-1980 | 1981-1990 | 1991-2000 | 2001-2010 | 2011-2018 |
|---|---|---|---|---|---|---|
| Spain | 4 | 3 | 0 | 2 | 2 | 2 |
| Italy | 2 | 4 | 5 | 1 | 1 | 0 |
| Ex USSR and Yugoslavian | 7 | 2 | 4 | 3 | 2 | 1 |
| Other | 0 | 1 | 1 | 4 | 6 | 5 |

Fig.2. Appearances to Final Four per geographic area

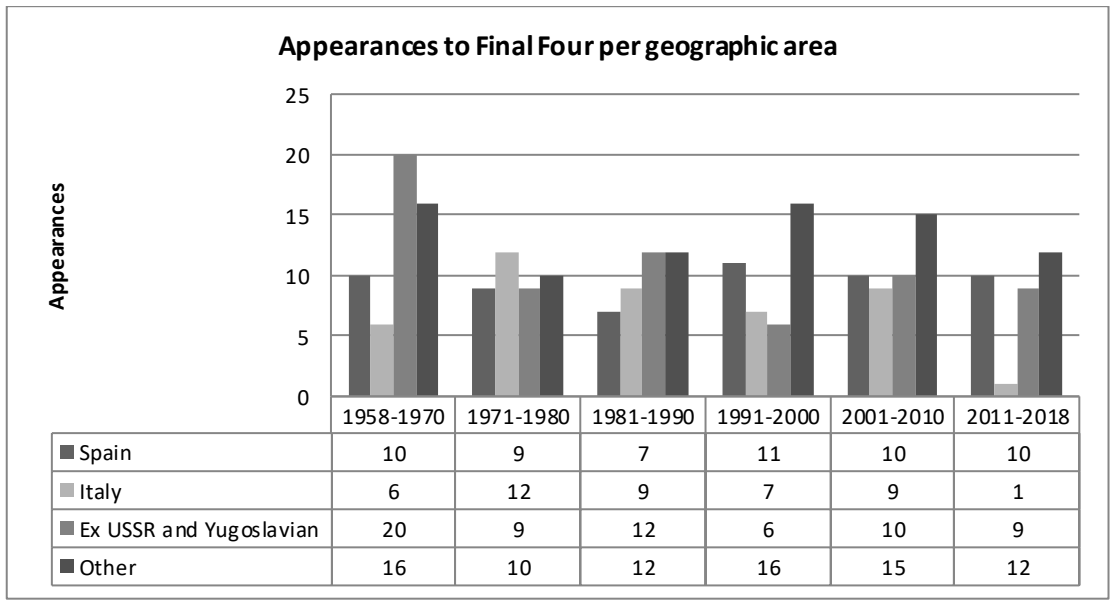

| | 1958-1970 | 1971-1980 | 1981-1990 | 1991-2000 | 2001-2010 | 2011-2018 |
|---|---|---|---|---|---|---|
| Spain | 10 | 9 | 7 | 11 | 10 | 10 |
| Italy | 6 | 12 | 9 | 7 | 9 | 1 |
| Ex USSR and Yugoslavian | 20 | 9 | 12 | 6 | 10 | 9 |
| Other | 16 | 10 | 12 | 16 | 15 | 12 |

To test formally if there are significant differences to the appearances and to the titles per geographic area, we perform Friedman tests with titles (or appearances) per geographic area as treatments and time periods as blocks (a blocking factor is a source of variability which is not of primary interest). We want to check for significant differences to the titles and appearances per geographic area. Note that we want to overall check the titles and appearances and not the trend, and we consider the time periods as blocks in order to reduce their effect to the variability of titles and appearances respectively. The non-parametric Friedman test is used in order not to have distributional assumptions, because normality assumption (data to follow normal distribution) does not seem very likely. Details about the test can be found in every book of non-parametric statistics such as that of (Hollander and Wolfe, 1999).

The null hypothesis ($H_0$) is that apart of the effect of time period (blocks) there is no difference in titles (or appearances) are even between the considered regions. The level of significance is considered at 5% (0.05). To reject the null hypothesis, the p-value should be less than 0.05.

| Friedman Test | | | |
|---|---|---|---|
| | *Statistic* | *df* | *p-value* |
| Appearances | 6.5789 | 3 | 0.0866 |
| Titles | 0.7627 | 3 | 0.8584 |

From the application of the test we do not have enough evidence to suppose significant differences between the performance of geographic areas in appearances and titles. However there are trends which have been described graphically and discussed previously.

### 3.2 Dominance of the Champion

Next, is examined the dominance of the champion to its opponents and is measured in terms of scoring points. The considered data are the points per game for and against the champions after the quarterfinals because the potential existence of weak teams in earlier rounds may lead to instability of point performance. In Fig.3 there are displayed the Points per Game (PPG) of the champion team and of their opponents, while Fig.4 displays the Point difference as % of the points of the opponents of the champion team. This is considered as a metric of the dominance of the champion team against its opponents in terms of scoring.

Fig.3. PPG for and against the champion team

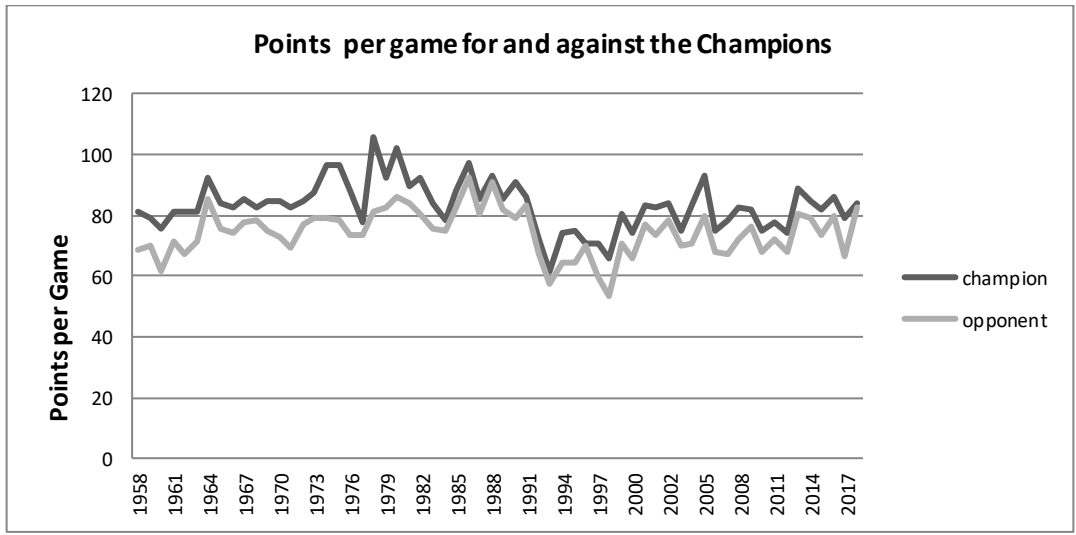

Fig.4. Point difference as % of the opponents of the champion team

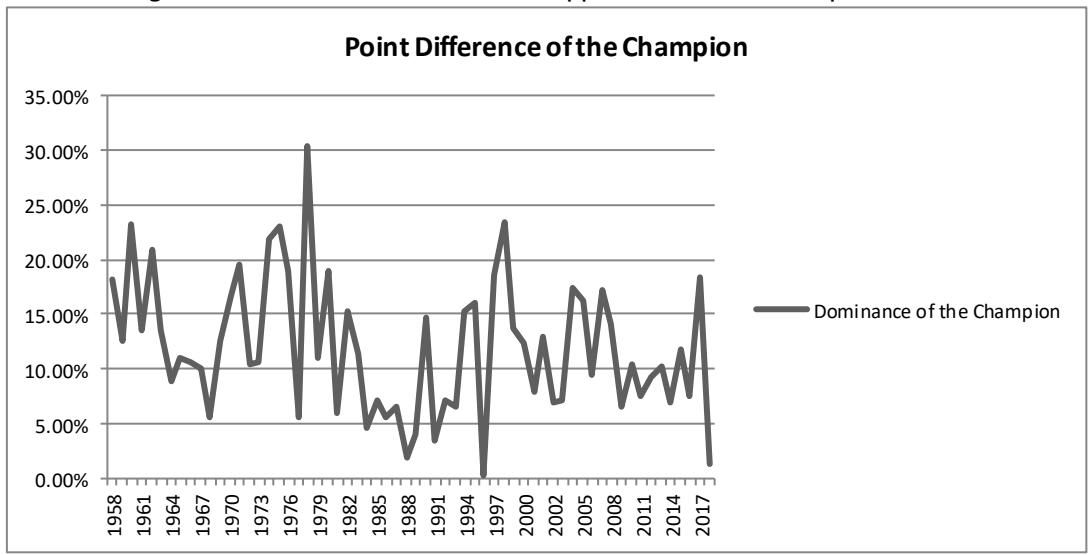

The above graph displays the points scored by the champion minus the points scored by opponents on average, as percentage of opponent points. This could show in a sense how dominant was a champion.

Only in six seasons the champions scored more than 20% of their opponent points with Real Madrid to be the only team which scored more than 30% of their opponents' points in 1978. Extreme cases like this should be examined in more detail. For example this team scored only 75 points in the final.

To better track the change of the game over time, we calculate the average points of the champions and opponents in every decade and the average points per team and we draw their evolution across time in Table 2 and Fig.5.

Table 2. Euroleague For and Against Points for Champion after Quarterfinals on average

| Time Period | Average Points per Time Period | | |
|---|---|---|---|
| | champion | opponent | Points per Team |
| **1958-1970** | 82.61 | 72.90 | 77.75 |
| **1971- 1980** | 91.20 | 77.88 | 84.54 |
| **1981-1990** | 88.27 | 81.92 | 85.09 |
| **1991-2000** | 72.92 | 65.61 | 69.27 |
| **2001-2010** | 82.94 | 74.38 | 78.66 |
| **2011-2018** | 81.86 | 75.18 | 78.52 |

Fig.5. Evolution of Average Points per team

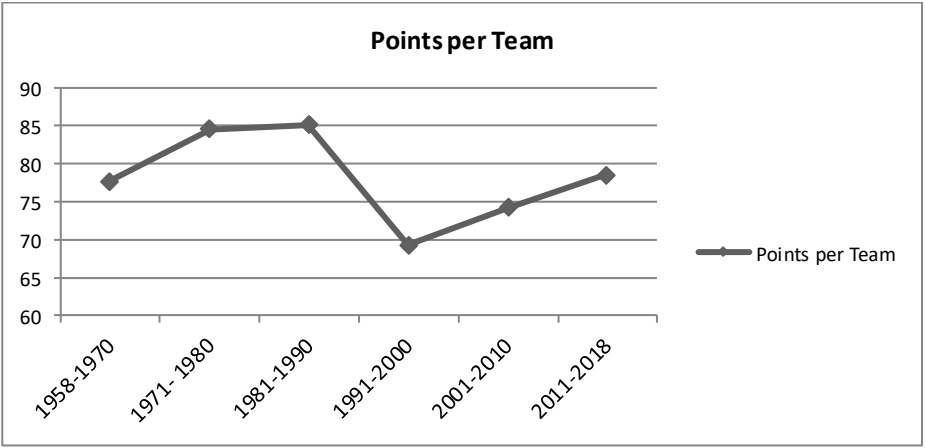

From the above graph we can see the changes of the mentality of the game across time. From 1958 to 1990 there was an upward trend in scoring, with a sudden drop in 90s, something which indicates a significant change in game mentality, and a return to the levels of 1958-1970 period.

Finally, the Fig.6 displays the decade average of the difference of points scored by champions minus the points scored by opponents as percentage of the opponent points. It's an indication of the dominance of the champions of every decade. On average, the champions were more dominant in 60s and 70s, but in the 80s they scored only 8% more than their opponents, a clear sign that the competition was more intense in this decade. We also notice that the competitiveness of the last years (2010-2018) tends to similar levels.

Fig.6. Evolution of Point difference as % of the opponents of the champion team

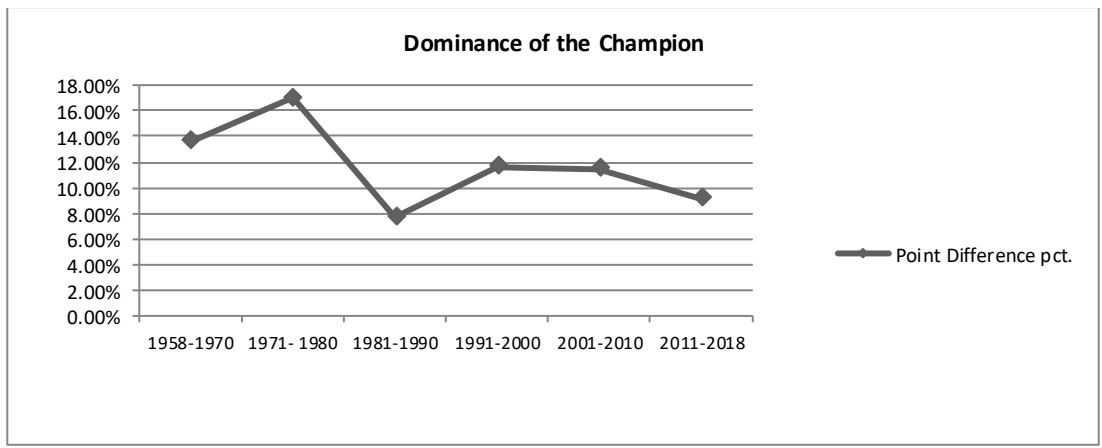

### 3.3 Analyze Scoring Pluralism in the Finals: Evolution of the style of the game

In this subsection we want to follow the evolution of the game as pictured in finals. We use raw data which are the first scorers and the team points of the finals since 1958. It is commonly assumed that in the runner-up team there is a more dominant scorer, in terms of first scorer points as % of team points. To test this hypothesis we perform a Wilcoxon rank sum test for pairs of observations for data from all finals since 1958. The null hypothesis ($H_0$) of the test is that the differences between the pairs follows a symmetric distribution around zero. This is a non parametric test and through its application we avoid the distributional assumption of normality of the data. A detailed description of the test can be found in (Hollander and Wolfe, 1999). The test suggests that there is no reason to assume that in a specific year is more probable the runner-up team to have a more dominant player in the scoring in final.

| Wilcoxon Signed Rank Test for paired Samples ||
|---|---|
| *Statistic* | *p-value* |
| 824 | 0.1494 |

Additionally, we explore whether first scorer in terms of % of team points appears randomly or is more probable to appear in sequences either from the champion or from the runner-up team. This can be achieved through the application of a runs test in the difference of first scorer points as % of team points between the two teams and we generate from this variable a sequence of + signs (if the variable is larger than a threshold) and – signs (if the variable is smaller than a threshold). A run of a sequence is defined as a series consisting of adjacent equal elements. We are testing the null hypothesis ($H_0$) that each element in the sequence is independently drawn from the same distribution. The threshold in our case is set to zero. A description of the test can be found in (Gibbons and Chakraborti, 2003) and its implementation performed via the *randtests* package of R (Caeiro and Mateus, 2014).

Through the application of the test, we can decide if over and under zero values are random. There is no sign of non-randomness for this variable, so we can assume that the first scorer appears randomly from the champion or from the runner-up team.

| Runs Test |||||
|---|---|---|---|---|
| *Statistic* | *Observations>0* | *Observations<0* | *Runs* | *p-value* |
| ~0 | 24 | 40 | 31 | ~1 |

The above tests are for the whole time period and they don't reveal anything about the evolution of the game. The rest of this section examines the evolution of the game and the statistical tests are adjusted accordingly.

At first, we present graphically the 10 year moving average of the points scored by the first scorer in final as % of team points in Fig.7. It is displayed the 10 year moving average for decreasing the effect of extreme cases and is easier to follow the trend of the game.

Fig.7. Moving average (10-year) of the points scored by the first scorer in final as % of team points

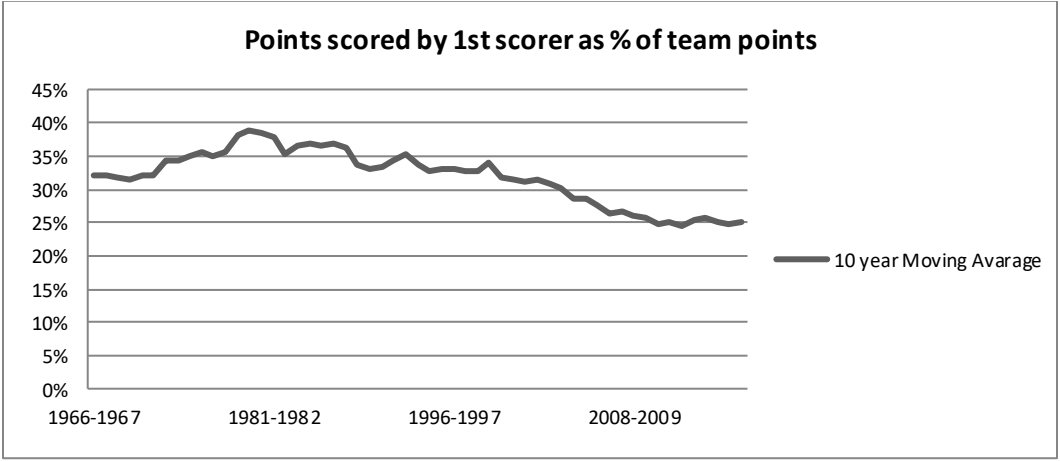

We observe that the game has been transformed over time from finals with offences which are based on top scorers to finals with more pluralism. From the beginning until the 80s the trend was the one player star in scoring, but since then, there was a slow but continuing turn to games based on pluralism.

At the next step, in Fig. 8 we present graphically the 10 year moving average of the difference between the points of first scorer as % of team points for champion team minus the same metric of runner-up team.

Fig.8. Moving average (10-year) of the difference of points of first scorers as % of their team points

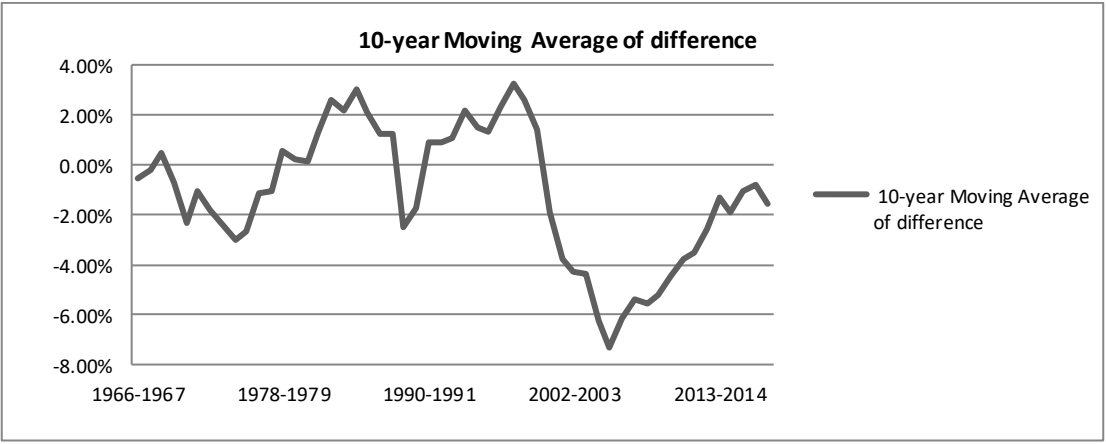

From 1998 there is a downward trend until 2009 and a new cycle begins after 2009 and evolves but in a lower level than the past. After 2001 there is no single year where the champion team has a more dominant scorer than the runner-up team in terms of 10 year moving averages.

We make an assumption that there is a structural break in this variable and is very important to specify the time when it happened because is a clue that the game has changed at this moment. To achieve this, is performed a Zivot-Andrew test (Zivot and Andrews, 1992) to test for the existence of a structural break (null hypothesis $H_0$) against the hypothesis of nonstationarity.

| Andrew-Zivot Test* | | |
|---|---|---|
| *Statistic* | *p-value* | *Potential Break* |
| -4.3499 | >0.1 | 1997-1998 final |

*We assume both level and linear trend and 5 lags*

The existence of a structural break is in favour compared to non-stationarity. Moreover, it is important to specify when the structural break occurs. The potential structural break occurs in 1997-1998 final. The history of Euroleague can be break into 2 periods: before and after 1998, let's say after 1998 is the modern period of Euroleague. For this reason, we perform again the Wilcoxon sign rank test and the Runs test for the modern period of Euroleague. In the modern period of Euroleague, we can assume that there is a more dominant scorer in the runner-up team, but we cannot predict this fact for a specific year.

| Wilcoxon Signed Rank Test for paired Samples | |
|---|---|
| *Statistic* | *p-value* |
| 42.5 | 0.02056 |

| Runs Test | | | | |
|---|---|---|---|---|
| *Statistic* | *Observations>0* | *Observations<0* | *Runs* | *p-value* |
| 1.5607 | 5 | 12 | 11 | 0.1186 |

## 3.4 Pace in the Finals: The concept of possessions

In this subsection we include to our analysis the central concept of possessions (Kubatko et al, 2007). Larger number of possessions displays a quicker pace of a game and the intension is to track the evolution of the game.

We assume that both teams have the same number of possessions, but there is no unique formula for the calculation of exact possessions in a game. For this reason, there are considered two formulas and we average them in order to approximate more accurately the actual possessions. The used formulas are the possessions lost (1) and the possessions gained (2) respectively:

$$POSS_t = FGA_t + \lambda \times FTA_t - OREB_t + TO_t \qquad (1)$$

$$POSS_t = FGM_t + \lambda \times FTM_t + DREB_o + TO_t \qquad (2)$$

After the calculation of the positions, we perform a line graph for the 5 year moving average of the possessions in order to track their evolution (Fig.9). There is a downward trend and stability in low game pace in the 90s, but from the beginning of the millennium there is a growing trend in game pace and from 2002 only four times there were fewer than 70 positions.

Fig.9. Evolution of possessions as 5-year moving average

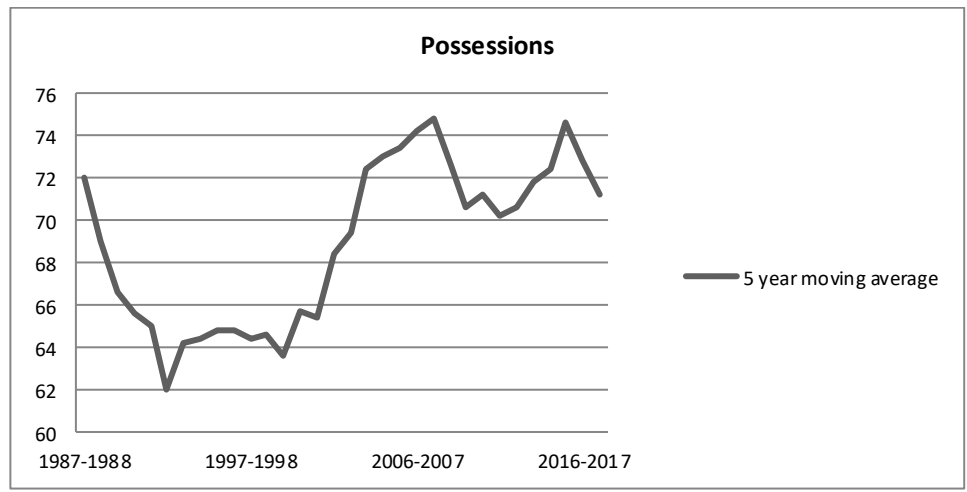

Considering a break at 1997-1998 we perform a Mann-Whitney test for the equality of possessions before and after 1998. This test is a non parametric equivalent of a t-test for comparing the means of 2 groups when the data do not follow Normal distribution. Details can be found on (Hollander and Wolfe, 1999).

| Possessions | |
|---|---|
| *Before 1998* | *After 1998* |
| 66.25 | 71.33 |
| **Mann-Whitney Test for possessions** | |
| *Statistic* | *p-value* |
| 84 | 0.01028 |

We detect a significant difference in possessions before and after 1998 finals at the 5% significance level. This result is in accordance with our assumption that the game has changed after 1998 final and supports the assumption that the triumph of Zalgiris in 1999 was the start of the change.

### 3.5  Correlation of First Scorer with Team Performance

Another interesting question is whether the existence of a first scorer of a tournament is correlated with the performance of the team. The most popular opinion is that first scorers belong to weak teams which do not have offensive many good offensive players. The other opinion is that a very gifted scorer can affect the performance of the team positively and relies to the coach to keep the balance of the team. We have the first scorers of the tournament after 1992 and we measure the strength of the link between their scoring performances (PPG) with the success of their team in the season using the Pearson (r) and Spearman ($\rho$) correlation coefficients (Hollander and Wolfe, 1999; Best and Roberts, 1975). The Spearman coefficient is non parametric and correlates the ranks of the variables and assesses monotonic relationships between them (instead of linear relationships which are assessed from Pearson correlation). Except from the coefficient, we perform a statistical test for testing the null hypothesis that the coefficient (either r or $\rho$) is zero, thus there is not significant correlation between the variables. We assign values for the performance of the teams: 1 for regular season, 2 for Top 16, 3 for quarterfinals, 4 for final four, 4.5 if the team was runner-up and 5 if the team won the trophy. Table 3 displays the first scorer, the team position and the assigned values of the position.

|  | | $H_0: \rho = 0 \; or \; r = 0$ | |
|---|---|---|---|
|  | **Coefficient** | **Statistic** | **p-value** |
| **Pearson Correlation** | -0.4002 | -2.2271 | 0.03481 |
| **Spearman Correlation** | -0.3925 | 5088.032 | 0.03886 |

Both Pearson and Spearman correlation coefficients indicate that there is a significant negative relationship between the first scorer and the performance of the team. This finding rather favors the first opinion where the first scorers are rarely parts of top teams (exception of Nando De Colo in 2015-2016 confirms the general rule).

Table 3. First scorer of the tournament, team position and assigned values of the position

| Season | Player | PPG | Team | Performance | Assigned Score |
|---|---|---|---|---|---|
| **1991-1992** | Nikos Galis | 32.3 | Aris | Regular season | 1 |
| **1992-1993** | Zdravko Radulović | 23.9 | Cibona | Regular season | 1 |
| **1993-1994** | Nikos Galis | 23.8 | Panathinaikos | 3rd place | 4 |
| **1994-1995** | Sašha Danilović | 22.1 | Buckler Bologna | Quarterfinals | 3 |
| **1995-1996** | Joe Arlauckas | 26.4 | Real Madrid | 4th place | 4 |
| **1996-1997** | Carlton Myers | 22.9 | Teamsystem Bologna | Quarterfinals | 3 |
| **1997-1998** | Peja Stojaković | 20.9 | PAOK | Top 16 | 2 |
| **1998-1999** | İbrahim Kutluay | 21.4 | Fenerbahçe | Top 16 | 2 |
| **1999-2000** | Miljan Goljović | 20.2 | Pivovarna Laško | Regular season | 1 |
| **2000-2001 (FIBA)** | Miroslav Berić | 23.3 | Partizan | Top 16 | 2 |
| **2000-2001 (Euroleague)** | Alphonso Ford | 26 | Peristeri | Top 16 | 2 |
| **2001-2002** | Alphonso Ford | 24.8 | Olympiacos | Top 16 | 2 |
| **2002-2003** | Miloš Vujanić | 25.8 | Partizan | Regular season | 1 |
| **2003-2004** | Lynn Greer | 25.1 | Śląsk Wrocław | Regular season | 1 |
| **2004-2005** | Charles Smith | 20.7 | Scavolini Pesaro | Quarterfinals | 3 |
| **2005-2006** | Drew Nicholas | 18.5 | Benetton Treviso | Top 16 | 2 |
| **2006-2007** | Igor Rakočević | 16.2 | Tau Cerámica | 4th place | 4 |
| **2007-2008** | Marc Salyers | 21.8 | Roanne | Regular season | 1 |
| **2008-2009** | Igor Rakočević | 18 | Tau Cerámica | Quarterfinals | 3 |
| **2009-2010** | Linas Kleiza | 17.1 | Olympiacos | 2nd place | 4.5 |
| **2010-2011** | Igor Rakočević | 17.2 | Efes Pilsen | Top 16 | 2 |
| **2011-2012** | Bo McCalebb | 16.9 | Montepaschi Siena | Quarterfinals | 3 |
| **2012-2013** | Bobby Brown | 18.8 | Montepaschi Siena | Top 16 | 2 |

| 2013-2014 | Keith Langford | 17.6 | EA7 Milano | Quarterfinals | 3 |
| 2014-2015 | Taylor Rochestie | 18.9 | Nizhny Novgorod | Top 16 | 2 |
| 2015-2016 | Nando de Colo | 18.9 | CSKA Moscow | Winner | 5 |
| 2016-2017 | Keith Langford | 21.8 | UNICS | Regular season | 1 |
| 2017-2018 | Alexey Shved | 21.8 | Khimki | Quarterfinals | 3 |

### 3.6 Unexpected Outcomes in the Final Fours

In this section it is examined the unexpected of the Final-Fours in terms of outcomes based on previous attempts with the use of Binomial Distribution. Can we make the assumption that each final four is an experiment with each team to have the same probabilities of winning the tournament (25%)?

To answer this question, we consider each final four as an experiment and teams are considered as independent random variables. Each experiment can be described by the binomial distribution and the whole situation with multinomial distribution (Forbes et al, 2011) which is a generalization of binomial distribution and describes n trials. There is performed a multinomial goodness of fit test and to strengthen the results a binomial test for each team, in order to decide if there is any significant difference from binomial distribution.

| **Multinomial Testing** | | *p-value* |
|---|---|---|
| | | 0.54499±0.001575 |
| **Binomial Testing*** | | |
| | | *p-value* |
| Cibona | *2 attempts - 2 trophies* | 0.0625 |
| Jugoplastica | *4 attempts - 3 trophies* | 0.05078 |
| ASK Riga | *4 attempts - 3 trophies* | 0.05078 |
| Panathinaikos | *12 attempts - 6 trophies* | 0.08608 |

*Only cases with p-value<0.1*

Table on the appendix displays the final four teams, the expected titles according to Binomial distribution, the observed values and their difference. Indeed there is no evidence that there are significant discrepancies from the binomial distribution at the 5% level of significance.

In the modern period of Euroleague (1999-2018), again there is no evidence of significant discrepancy from the multinomial distribution, but the case of Panathinaikos could be seen as an exception, with significant larger success rate than the expected.

| Multinomial Testing | | *p-value* |
|---|---|---|
| | | 0.68173±0.001473 |
| **Binomial Testing*** | | |
| | | *p-value* |
| Panathinaikos | 8 attempts - 5 trophies | 0.02730 |

*\*Only cases with p-value<0.1*

## 4 Summary and Conclusions

To sum up, in this paper is made an attempt to address questions related to historical evolution of Euroleague using statistical analysis to draw conclusions. One main problem is the lack of plenty available statistical data from the beginning of the competition. This paper demonstrates that by applying suitable statistical designs we can draw interesting conclusions even with limited data. Firstly, is made a brief exploration of the historical evolution of the Euroleague and the tracking statistics.

Then, some questions are answered and some conclusions are drawn which are briefly the following:

Although overall there is no difference in success between more traditional powers such as Italy, Spain and ex USSR and Yugoslavian countries and other countries, there is a clear trend of other countries to expand their presence (in terms of titles and final four appearances) in the tournament after the 90s. In terms of scoring, there was an upward trend from the beginning of the competition, with a sudden drop in 90s, something which indicates a significant change in game mentality in terms of defence and/or game pace. The champions were more dominant in 60s and 70s, but in the 80s they scored only 8% more than their opponent, which indicates that the competition was more intense in this decade. The last years (2010-2018), the competitiveness of the tournament tends to similar levels.

There is a popular belief that the first scorer in the majority of cases come from the runner-up team. However, there is no reason to assume that in a specific year is more probable the runner-up team to have a more dominant player in the scoring in the final. In the modern period of Euroleague (after 1998), we can assume that there is a more dominant scorer in the runner-up team, but we cannot predict this fact for a specific year. According to the game evolution in finals, we observe that the game has been transformed from finals with offences which are based on top scorers to finals with more pluralism. From the beginning until the 80s the trend was the one player star in scoring, but since then, there was a slow but continuing turn to games based on pluralism. Moreover, we detect a significant difference in possessions before and after 1998 finals, which is in accordance with our assumption that the game has changed after 1998 final and supports the assumption that the triumph of Zalgiris in 1999 was the start of the change. Furthermore, it is found a significant negative relationship between the first scorer and the performance of his team. This finding favors the opinion that the first scorers are rarely parts of top teams. Finally, there is no evidence to reject the hypothesis that in a final four there are equal chances of winning overall. In modern era, again the hypothesis of the final four as a random experiment is not rejected, however in the case of Panathinaikos we observe significantly higher success rate than the expected.

# Appendix

Table. Euroleague Final Four Teams

| Year | Winner | Runner-Up | 3rd Place | 4th Place |
|---|---|---|---|---|
| 1958 | ASK Riga | Academic | Honved | Real Madrid |
| 1958-1959 | ASK Riga | Academic | Lech Poznan | OKK Beograd |
| 1959-1960 | ASK Riga | Dinamo Tbilisi | Pologna Warzawa | Slovan Orbis Praha |
| 1960-1961 | CSKA Moscow | ASK Riga | Real Madrid | Steaua Bucarest |
| 1961-1962 | Dinamo Tbilisi | Real Madrid | ASK Olimpija | CSKA Moscow |
| 1962-1963 | CSKA Moscow | Real Madrid | Dinamo Tbilisi | Spartak Brno |
| 1963-1964 | Real Madrid | Spartak Brno | OKK Beograd | Simmenthal Milano |
| 1964-1965 | Real Madrid | CSKA Moscow | Ignis Varese | OKK Beograd |
| 1965-1966 | Simmenthal Milano | Slavia Praha | CSKA Moscow | AEK |
| 1966-1967 | Real Madrid | Simmenthal Milano | ASK Olimpija | Slavia Praha |
| 1967-1968 | Real Madrid | Spartak Brno | Simmenthal Milano | Zadar |
| 1968-1969 | CSKA Moscow | Real Madrid | Spartak Brno | Standard Liege |
| 1969-1970 | Ignis Varese | CSKA Moscow | Real Madrid | Slavia Praha |
| 1970-1971 | CSKA Moscow | Ignis Varese | Real Madrid | Slavia Praha |
| 1971-1972 | Ignis Varese | Jugoplastica[1] | Real Madrid | Panathinaikos |
| 1972-1973 | Ignis Varese | CSKA Moscow | Crvena Zvezda | Simmenthal Milano |
| 1973-1974 | Real Madrid | Ignis Varese | Berck | Radniski Belgrade |
| 1974-1975 | Ignis Varese | Real Madrid | Berck | Zadar |
| 1975-1976 | Mobilgirgi Varese | Real Madrid | ASVEL | Forst Cantù |
| 1976-1977 | Maccabi Tel Aviv | Mobilgirgi Varese | CSKA Moscow | Real Madrid |
| 1977-1978 | Real Madrid | Mobilgirgi Varese | ASVEL | Maccabi Tel Aviv |
| 1978-1979 | Bosna | Emerson Varese | Maccabi Tel Aviv | Real Madrid |
| 1979-1980 | Real Madrid | Maccabi Tel Aviv | Bosna | Sinudyne Bologna[3] |
| 1980-1981 | Maccabi Tel Aviv | Sinudyne Bologna[3] | Nashua EBBC | Bosna |
| 1981-1982 | Squibb Cantù | Maccabi Tel Aviv | Partizan | FC Barcelona |
| 1982-1983 | Ford Cantù | Billy Milano | Real Madrid | CSKA Moscow |
| 1983-1984 | Virtus Roma | FC Barcelona | Jollycolombani Cantù | Bosna |
| 1984-1985 | Cibona | Real Madrid | Maccabi Tel Aviv | CSKA Moscow |
| 1985-1986 | Cibona | Zalgiris | Simac Milano | Real Madrid |
| 1986-1987 | Tracer Milano | Maccabi Tel Aviv | Orthez | Zadar |
| 1987-1988 | Tracer Milano | Maccabi Tel Aviv | Partizan | Aris |
| 1988-1989 | Jugoplastica[1] | Maccabi Tel Aviv | Aris | FC Barcelona |
| 1989-1990 | Jugoplastica[1] | FC Barcelona | Limoges | Aris |
| 1990-1991 | POP 84[1] | FC Barcelona | Maccabi Tel Aviv | Scavolini Pezaro |
| 1991-1992 | Partizan | Joventut | Phillips Milano | Estudiantes |
| 1992-1993 | Limoges | Benneton Treviso | PAOK | Real Madrid |
| 1993-1994 | Joventut | Olympiacos | Panathinaikos | FC Barcelona |
| 1994-1995 | Real Madrid | Olympiacos | Panathinaikos | Limoges |
| 1995-1996 | Panathinaikos | FC Barcelona | CSKA Moscow | Real Madrid |
| 1996-1997 | Olympiacos | FC Barcelona | Smelt Olimpija | ASVEL |
| 1997-1998 | Kinder Bologna[3] | AEK | Benneton Treviso | Partizan |
| 1998-1999 | Zalgiris | Kinder Bologna[3] | Olympiacos | Teamsystem Bologna[4] |
| 1999-2000 | Panathinaikos | Maccabi Tel Aviv | Efes Pilsen | FC Barcelona |
| 2000-2001 (FIBA) | Kinder Bologna[3] | TAU Ceramica[2] | AEK | Paf Wennington Bologna[4] |
| 2000-2001 (Euroleague) | Maccabi Tel Aviv | Panathinaikos | Efes Pilsen | CSKA Moscow |
| 2001-2002 | Panathinaikos | Kinder Bologna[3] | Benneton Treviso | Maccabi Tel Aviv |
| 2002-2003 | FC Barcelona | Benneton Treviso | Montepaschi Siena | CSKA Moscow |
| 2003-2004 | Maccabi Tel Aviv | Skipper Bologna[4] | CSKA Moscow | Montepaschi Siena |
| 2004-2005 | Maccabi Tel Aviv | TAU Ceramica[2] | Panathinaikos | CSKA Moscow |
| 2005-2006 | CSKA Moscow | Maccabi Tel Aviv | TAU Ceramica[2] | FC Barcelona |
| 2006-2007 | Panathinaikos | CSKA Moscow | Unicaja Malaga | TAU Ceramica[2] |
| 2007-2008 | CSKA Moscow | Maccabi Tel Aviv | Montepaschi Siena | Real Madrid |
| 2008-2009 | Panathinaikos | CSKA Moscow | FC Barcelona | Olympiacos |
| 2009-2010 | FC Barcelona | Olympiacos | CSKA Moscow | Partizan |
| 2010-2011 | Panathinaikos | Maccabi Tel Aviv | Montepaschi Siena | Real Madrid |
| 2011-2012 | Olympiacos | CSKA Moscow | FC Barcelona | Panathinaikos |
| 2012-2013 | Olympiacos | Real Madrid | CSKA Moscow | FC Barcelona |
| 2013-2014 | Maccabi Tel Aviv | Real Madrid | FC Barcelona | CSKA Moscow |
| 2014-2015 | Real Madrid | Olympiacos | CSKA Moscow | Fenerbahce |
| 2015-2016 | CSKA Moscow | Fenerbahce | Lokomotiv Kuban | Laboral Kutxa[2] |
| 2016-2017 | Fenerbahce | Olympiacos | CSKA Moscow | Real Madrid |
| 2017-2018 | Real Madrid | Fenerbahce | Zalgiris | CSKA Moscow |

[1]*Croatia Split*

[2]*Club Deportivo Saski Baskonia, S.A.D.*

[3]*Virtus Pallacanestro Bologna*

[4] *Fortitudo Pallacanestro Bologna 103*

Table . Teams and appearances to final four

| Team | Winner | Appearances | Expected Titles | Observed Titles | Difference |
|---|---|---|---|---|---|
| Lokomotiv Kuban | 0 | 1 | 0.25 | 0 | -0.25 |
| Unicaja Malaga | 0 | 1 | 0.25 | 0 | -0.25 |
| PAOK | 0 | 1 | 0.25 | 0 | -0.25 |
| Estudiantes | 0 | 1 | 0.25 | 0 | -0.25 |
| Scavolini Pezaro | 0 | 1 | 0.25 | 0 | -0.25 |
| Orthez | 0 | 1 | 0.25 | 0 | -0.25 |
| Nashua EBBC | 0 | 1 | 0.25 | 0 | -0.25 |
| Radniski Belgrade | 0 | 1 | 0.25 | 0 | -0.25 |
| Crvena Zvezda | 0 | 1 | 0.25 | 0 | -0.25 |
| Standard Liege | 0 | 1 | 0.25 | 0 | -0.25 |
| Steaua Bucarest | 0 | 1 | 0.25 | 0 | -0.25 |
| Lech Poznan | 0 | 1 | 0.25 | 0 | -0.25 |
| Honved | 0 | 1 | 0.25 | 0 | -0.25 |
| Pologna Warzawa | 0 | 1 | 0.25 | 0 | -0.25 |
| Efes Pilsen | 0 | 2 | 0.5 | 0 | -0.5 |
| Berck | 0 | 2 | 0.5 | 0 | -0.5 |
| Zadar | 0 | 3 | 0.75 | 0 | -0.75 |
| Olimpija Ljubliana | 0 | 3 | 0.75 | 0 | -0.75 |
| OKK Beograd | 0 | 3 | 0.75 | 0 | -0.75 |
| ASVEL | 0 | 3 | 0.75 | 0 | -0.75 |
| Aris | 0 | 3 | 0.75 | 0 | -0.75 |
| Montepaschi Siena | 0 | 4 | 1 | 0 | -1 |
| Fortitudo Bologna | 0 | 3 | 0.75 | 0 | -0.75 |
| AEK | 0 | 3 | 0.75 | 0 | -0.75 |
| Praha | 0 | 5 | 1.25 | 0 | -1.25 |
| Baskonia | 0 | 5 | 1.25 | 0 | -1.25 |
| Treviso | 0 | 4 | 1 | 0 | -1 |
| Brno | 0 | 4 | 1 | 0 | -1 |
| Academic | 0 | 2 | 0.5 | 0 | -0.5 |
| Limoges | 1 | 3 | 0.75 | 1 | 0.25 |
| Partizan | 1 | 5 | 1.25 | 1 | -0.25 |
| Virtus Roma | 1 | 1 | 0.25 | 1 | 0.75 |
| Bosna | 1 | 4 | 1 | 1 | 0 |
| Zalgiris | 1 | 3 | 0.75 | 1 | 0.25 |
| Joventut Badalona | 1 | 2 | 0.5 | 1 | 0.5 |
| Dinamo Tbilisi | 1 | 3 | 0.75 | 1 | 0.25 |
| Fenerbahce | 1 | 4 | 1 | 1 | 0 |
| Cibona | 2 | 2 | 0.5 | 2 | 1.5 |
| Cantu | 2 | 4 | 1 | 2 | 1 |
| Virtus Bologna | 2 | 6 | 1.5 | 2 | 0.5 |
| FC Barcelona | 2 | 16 | 4 | 2 | -2 |
| Split | 3 | 4 | 1 | 3 | 2 |
| ASK Riga (Latvia) | 3 | 4 | 1 | 3 | 2 |
| Olympia Milano | 3 | 10 | 2.5 | 3 | 0.5 |
| Olympiacos | 3 | 10 | 2.5 | 3 | 0.5 |
| Varese | 5 | 11 | 2.75 | 5 | 2.25 |
| Panathinaikos | 6 | 12 | 3 | 6 | 3 |
| Maccabi Tel Aviv | 6 | 20 | 5 | 6 | 1 |
| CSKA Moscow | 7 | 29 | 7.25 | 7 | -0.25 |
| Real Madrid | 10 | 32 | 8 | 10 | 2 |